\documentclass[prd,onecolumn, aps, superscriptaddress, preprintnumbers, floatfix, nofootinbib]{revtex4}

\usepackage{amssymb}
\usepackage{amsmath}
\usepackage{graphicx}
\usepackage{subfig}
\begin{document}

\title{On general features of  warm dark matter with reduced relativistic gas}

\author{W. S. Hip\'olito-Ricaldi
\footnote{E-mail:wiliam.ricaldi@ufes.br}}
\affiliation{Universidade Federal do Esp\'{\i}rito Santo, Departamento de Ci\^encias Naturais, Rodovia BR 101 Norte,
km. 60, S\~ao Mateus, ES, Brazil.}
\affiliation{Facultad de Ciencias, Universidad Nacional de Ingenier\'{\i}a, Av T\'upac Amaru. 210,Rimac, Lima, Per\'u.}

\author{R.F. vom Marttens\footnote{E-mail:rodrigovonmarttens@gmail.com}}
\affiliation{N\'ucleo Cosmo-ufes \& Departamento de F\'isica, CCE, Universidade Federal do Esp\'irito Santo, 29075-910, Vit\'oria-ES, Brazil.}

\author{J.C.Fabris\footnote{E-mail:fabris@pq.cnpq.br}}
\affiliation{N\'ucleo Cosmo-ufes \& Departamento de F\'isica, CCE, Universidade Federal do Esp\'irito Santo, 29075-910, Vit\'oria-ES, Brazil.}
\affiliation{National Research Nuclear University “MEPhI”, Kashirskoe sh. 31, Moscow 115409, Russia.}

\author{I.L.Shapiro\footnote{
E-mail:shapiro@fisica.ufjf.br}}
\affiliation{Universidade  Federal  de  Juiz  de  Fora, Departamento  de  F\'{\i}sica  –  ICE,  Juiz  de  Fora, CEP 36036-330,  MG, Brazil.}
\affiliation{Tomsk State Pedagogical University, 634041, Tomsk, Russia.}
\affiliation{Tomsk State University, 634050, Tomsk, Russia.}

\author{L. Casarini\footnote{
E-mail:casarini.astro@gmail.com}}
\affiliation{International Institute of Physics (IIP), Universidade Federal do Rio Grande do Norte (UFRN) CP 1613, 59078-970 Natal-RN, Brazil.}
\affiliation{Institute  of  Theoretical  Astrophysics, University  of  Oslo,  0315  Oslo,  Norway.}
\date{\today}

\begin{abstract}
Reduced Relativistic Gas (RRG) is a useful approach to describe the
warm dark matter (WDM) or the warmness of baryonic matter in the
approximation when the interaction between the particles is irrelevant.
The use of
Maxwell distribution leads to the complicated equation of state of
the J\"{u}ttner model of relativistic ideal gas. The RRG enables one
to reproduce the same physical situation but in a much simpler form. For
this reason RRG can be a useful tool for the theories with some sort
of a ``new Physics''. On the other hand, even without the qualitatively
new physical implementations, the RRG can be useful to describe the
general features of WDM in a model-independent way. In this sense one
can see, in particular, to which extent the cosmological manifestations
of WDM may be dependent on its Particle Physics background. In the
present work RRG is used as a complementary approach to derive the
main observational exponents for the WDM in a model-independent way.
The only assumption concerns a non-negligible velocity $v$ for dark
matter particles which is parameterized by the warmness parameter
$b$. The relatively high values of  $b$ ( $b^2\gtrsim 10^{-6}$) 
erase the radiation (photons and neutrinos) dominated
epoch and cause an early warm matter domination after inflation.
Furthermore, RRG approach enables one to quantify the  lack of
power in linear matter spectrum at small scales and in particular,
reproduces the relative transfer function commonly used in context
of WDM with  accuracy of $\lesssim 1\%$. A warmness with
$b^2\lesssim 10^{-6}$ (equivalent to $v\lesssim 300 km/s$)
does not alter significantly the CMB power spectrum and is in
agreement with the background observational tests.
\end{abstract}

\maketitle

\section{Introduction}
\label{sec:1}
In the last decades cosmological observations provided numerous
evidence for the two dark components nominated dark matter (DM)
and dark energy (DE), which are responsible for $\sim 96\%$ of the
content of the universe. In particular, the confirmation of existence
of these two dark components come from the measurements of the
luminosity redshift of type Ia supernova \cite{Riess:1998cb,Perlmutter:1998np}, baryon acoustic
oscillations \cite{Tegmark:2003uf}, anisotropies of the cosmic microwave
background (CMB)\cite{Aghanim:2015xee,Ade:2015xua} and  other observations
\cite{Moresco:2012jh}.  The standard interpretation suggests that DE is
necessary to accelerate the expansion of the universe. On the other
hand the DM has non-baryonic nature and is important, in particular,
to describe the formation of cosmic structure. The standard cosmology,
$\Lambda$CDM model, assumes that the DE is a cosmological
constant, and regards DM as a non-relativistic matter with  negligible
pressure (cold dark matter). $\Lambda$CDM  provides an excellent
agreement with most of the data (see, e.g., \cite{Bergstrom:2000pn} for
a general review), however this agreement is not perfect due to the
tensions with part of the  observational data (see for example
\cite{Buchert:2015wwr}).  In part due to these difficulties, some alternative
models have been proposed and studied as possible DE and DM
candidates (see for example \cite{Caldwell:1997ii,Liddle:1998xm,Peebles:2002gy,Sahni:2002dx,Lue:2005ya}). Let us note that
some of these alternative models aim to describe fluids that replace
both DM and DE (see for example \cite{Kamenshchik:2001cp,Fabris:2001tm,Bilic:2001cg,HipolitoRicaldi:2009je,Zimdahl:2011mb,Marttens:2017njo}) or
describe interaction between DE and DM \cite{Billyard:2000bh,Amendola:1999er,Honorez:2010rr,Zimdahl:2013yak,Wang:2016lxa,Funo:2014poa,Marttens:2014yja}.

Some of the mentioned $\Lambda$CDM difficulties are related
with the choice of the cold dark matter (CDM)  paradigm
\cite{Warren:2005ey}.  For instance, at small scales the issues such
as \textit{the missing satellites} problem \cite{Klypin:1999uc},
\textit{core/cusp} problem \cite{deBlok:2009sp}, and the
\textit{Too big to fail} problem \cite{BoylanKolchin:2011de}, can be alleviated by
assuming that the DM is not completely cold. In contrast
to the CDM,  the  Hot Dark Matter (HDM) scenario implies that
the free streaming due to a thermal motion of particles is important
to suppress structure formation at small scales. Nevertheless this
scenario was ruled out and opens the space for the Warm Darm
Matter (WDM) scenario. The main feature of WDM models is
that thermal velocities of the DM particles are not so high as in
the HDM scenario and, on the other hand, not negligible like
in the CDM scenario. Typically, the WDM models assume that
it is composed by particles of mass about $keV$  instead of $GeV$
which is ``typical'' for CDM and $eV$ which is the standard case for
the HDM. The standard approach to explore the possible warmness
of DM and its consequences for structure formation are  based on
to solution of the Hierarchy Bolztmann equation, taking into
account the specific properties of the given WDM candidate
\cite{Bond:1983hb,Hannestad:2000gt,Viel:2005qj,Bode:2000gq,Barkana:2001gr,Ma:1995ey}
\footnote {One has to remember that equations for DM are
always coupled to the Bolztmann equations for other components
of the universe.}.  For example,  relation between
mass and warmness for each WDM candidate comes  from the particle
physics arguments. This is in fact very good, because the ultimate
knowledge of the DM nature may be achieved only within the
particle physics and, more concrete, by means of laboratory
experiments.

Until the moment when the DM will be detected in the laboratory
experiments, one can always assume that the properties of DM
derived within a particle physics models may be violated by some
the qualitatively new scenarios for the DM, which can be never
ruled out completely \cite{Bergstrom:2000pn}. From this perspective, it
is useful to develop also model-independent approaches to
investigate the cosmological features of a WDM. In the present
work we will explore the consequences and impacts of warmness
in the process of structure formation and CMB anisotropies, but
using a model-independent approach which is based on the RRG 
approximation.
The RRG is a model of ideal gas of relativistic particles, which has
a very simple equation of state. This nice property is due to the main
assumption -- that the particles of the ideal gas have non-negligible
but equal thermal velocities. Regardless of this simplicity, the
model has long history which started in a glorious way. The RRG
equation of state was first introduced by A.D. Sakharov in order to
explore the acoustic features of Cosmic Microwave Background
(CMB) in the early universe \cite{sakharov1966initial}. Using this model
Sakharov predicted the existence of oscillations in CMB temperature
spectra long before its observational discovery (see
\cite{Grishchuk:2011wk} for the historical review).

Recently, RRG was reinvented in \cite{deBerredoPeixoto:2004ux}, where the
derivation of its equation of state was first presented explicitly.
The simplicity of the equation  of state is due to the assumption
that all particles of relativistic gas have equal kinetic energies,
i.e., equal velocities. Therefore RRG  is a reduced version of
well-known J\"{u}ttner model of relativistic ideal gas
\cite{Juttnerttner:1911,pauli1981theory}. A comparison
between the equations of state of the relativistic ideal gas and
RRG shows that the difference does not exceed $2.5\%$ even
in the low-energy region \cite{deBerredoPeixoto:2004ux} and becomes
completely negligible at higher energied. Further considerations
have shown that RRG model enables one to achieve a simple and
reliable description of the matter warmness in cosmology. In
Ref.~\cite{Fabris:2008qg} RRG was used to decribe WDM
and its perturbations were compared with the Large Scale
Structure data. Furthermore, the general analytic solutions
for the several background  cosmological models involving
RRG were discussed in Ref.~\cite{Medeiros:2012ud}.

The RRG was successfully used in \cite{deBerredoPeixoto:2004ux,Fabris:2008qg}
as an interpolation between radiation and dark matter eras in
cosmology.  An upper bound on the warmness coming from RRG
\cite{Fabris:2008qg,Fabris:2011am} is very close  to the one obtained from
much more complicated analysis based on a complete WDM
treatment, based on the Boltzmann equation. This standard approach
requires specifying the nature of the particle physics candidate for the
WDM contents \cite{Bond:1983hb,Hannestad:2000gt,Viel:2005qj,Bode:2000gq,Barkana:2001gr,Ma:1995ey}, while the approach based on RRG
requires only one parameter, that is the warmness of DM. In this
sense RRG represents a really useful tool for exploring WDM
cosmology  without specifying a particular candidate for the WDM.
Such a model may be helpful for better understanding of the
model-dependence or independence of the cosmological features
of WDM.

The main goal of the present work is to take advantage of the RRG
and its analytical solutions for the background cosmology and apply
it to WDM, instead of considering full set of WDM hierarchical
Boltzmann equations. The RRG enables one to make greater part
of considerations analytically and hence provide better qualitative
understanding of the results. In this way we consider the formation
of large-scale structure and the problem of CMB anisotropy in a
model-independent manner.
Following
\cite{Fabris:2008qg}, we shall establish the bounds for the thermal
velocities of the WDM particles in a general way. With this
objective in mind we consider the model of the spatially flat
Friedmann-Roberston-Walker universe filled by radiation \footnote{Of course,
WDM has a radiation behavior in early epochs but their physical processes are different than photons or neutrinos.
For this reason we will diferenciate in all paper WDM in early stages from ``standard``  radiation (photons and neutrinos).} and
RRG, representing WDM. Gravity is described by the general
relativity, with the cosmological constant representing DE.
We shall refer to this model as to $\Lambda$WDM.
All the perturbative treatments will be performed at the linear
order only, and using the normalization with the scale factor at
present $a_0=1$. With these notations, the WDM space of
parameters is reduced by using the most recent data from
SNIa, $H(z)$ and BAO.

The paper is organized as follows. Sec.~{\ref{sec:2} describes
the dynamics of the WDM in the framework of RRG, both at
the background and perturbative levels. It is shown that high
level of warmness may erase ''standard" radiation era from the cosmic
history. Starting from this point one can establish an upper
bound for the velocity of the RRG particles, which preserves
the ``standard`` primordial scenario for the universe. This bound is used
as a physical prior in the consideration of Sec.~{\ref{sec:3},
devoted to the statistical analysis using the background data.
In this framework we reduce the space of parameters for WDM
and use this reduced space in the consequent analysis. At the
next stage the CAMB code is modified and used to quantify
the relation between the DM warmness and  the total matter
density contrast, linear matter power spectrum and CMB power
spectrum. We show that the  RRG is capable to reproduce the
main feature of the WDM, i.e., the suppression of matter
over-densities at small scales. Furthermore, in Sec.~{\ref{sec:4} we discuss 
proprierties of thermal relics via RRG. Finally, Sec.~{\ref{sec:5} includes
discussions and conclusions.

\section{A description for a warm dark matter fluid}
\label{sec:2}

Let us start with the background notions. The reader can consult
\cite{deBerredoPeixoto:2004ux,Fabris:2008qg} or recent \cite{Reis:2017bjf} for
further details.

In the RRG approach WDM is treated  as an approximation of a
Maxwell-distributed ideal gas formed by massive particles. All
these particles have equal kinetic energies, or equal velocity
$\beta = v/c$ \cite{deBerredoPeixoto:2004ux} ($c$ is the light speed).
This leads to the following relation between WDM pressure
$p_w$ and WDM energy density $\rho_w$,

\begin{equation} \label{penergy}
p_w=\frac{\rho_w}{3}\left[1-\left(\frac{mc^2}{\epsilon}\right)^2\right],
\end{equation}
where $m$ is the WDM particle mass, and $\epsilon$ is the kinetic energy of each particle of the system which is given by,
\begin{equation} \label{epsilon}
\epsilon = \frac{mc^2}{\sqrt{1-\beta^2}}.
\end{equation}
Here $\rho_c$ is introduced as a notation for
the rest energy density, i.e.,  the energy density for the $v=0$ case.
Thus, $\rho_c=\rho_{c0}\, a^{-3}=n\,mc^{2}$, where $n$ is the
number  density $a=a(t)$ is the scale factor of the metric.
Using this relation, Eq.~(\ref{penergy}) can be cast into the form
\begin{equation} \label{pressure}
p_w=\dfrac{\rho_w}{3}\left[1-\left(\frac{\rho_c}{\rho_w}\right)^2\right],
\end{equation}
which can be regarded as equation of state (EoS) of the WDM fluid.

Using Eq.~(\ref{pressure}) in the energy conservation relation, the
solution for $\rho_w$ has the form
\begin{eqnarray}\label{density}
\rho_w(a) = \rho_{w0}\,a^{-3}\sqrt{\frac{1+b^{2}\,a^{-2}}{1+b^{2}}} \,, \qquad b=\frac{\beta}{\sqrt{1-\beta^2}} \,.
\end{eqnarray}
Thus, $b$ parameter measures velocity and warmness of the WDM particles
at present. In the limit $v\ll c$ we have $b \approx v/c$. Note also that for
$b=0$ the CDM case is recovered. Combining Eqs.~(\ref{pressure}) and
(\ref{density}), one can find \textit{a posteriori} state parameter,
\begin{eqnarray} \label{eos}
 w(a)=\frac{p_w}{\rho_w}=\frac{1}{3}-\frac{a^2}{3(a^2+b^2)} \,.
\end{eqnarray}
Here we called this term as a state parameter \textit{a posteriori} because
the "natural" EoS for the RRG description, given by Eq.~(\ref{pressure}),
implicitly depends on the scale factor and on the nowadays WDM energy
density. However after the integration of continuity equation it is possible
to write the state parameter that depends only on the scale factor. This form
will prove useful in the perturbative analysis.

In what follows we consider the  model with cosmological constant,
which does not agglomerate, WDM described by a RRG, baryons and
radiation. All of them are assumed being interacting gravitationally
and only photons and baryons also interacting via Thomson scattering
before recombination. In this situation Hubble rate takes the form
\begin{eqnarray}\label{hubble}
H^{2}=H_{0}^{2}\left(\Omega_{\Lambda0}+\dfrac{\Omega_{w0}}{a^{3}}\sqrt{\frac{1+b^{2}\,a^{-2}}{1+b^{2}}} + \dfrac{\Omega_{b0}}{a^{3}} + \dfrac{\Omega_{r0}}{a^{4}}\right).
\end{eqnarray}
In the last equation $\Omega_{x0}$ (with $x=\Lambda,w,b$ and $r$)
is the value of the DE, WDM, baryons and radiation density parameters
at present, while
$\Omega_{\Lambda0}=1-\Omega_{w0}-\Omega_{b0} -\Omega_{r0}$
since we deal with a spatially flat universe. It is easy to see that the
expressions (\ref{penergy}), (\ref{pressure}), (\ref{density}) and
(\ref{eos}) interpolate between the dust  at $b \to 0$ and radiation
at $b \to\infty$ cases. Because of this interpolation feature, RRG can be
used to investigate the cosmological consequences  of the transition
between epochs of radiation and  dust
\cite{sakharov1966initial,Grishchuk:2011wk,deBerredoPeixoto:2004ux}.

One can note that the WDM with EoS (\ref{pressure}) has a remarkable
consequence at early times, when RRG becomes very close to radiation.
This feature could cause an early warm matter domination and erase the
''standard" radiation dominated epoch. In order to ensure the existence of a ``standard" radiation
dominated era we must impose that in the very early universe the radiation
energy density is bigger than WDM energy density\footnote{We must emphasize that even though  in a primordial universe RRG behaves like radiation at 
background, this is
not true at perturbative level. On the other hand, processes involving  WDM radiation limit will be, in general, different that those 
involving the standard radiation (photons and neutrinos). The case where
WDM dominates even in early times deserves a more carefully study of
earlier processes like nucleosynthesis, reionization, reheating, etc and is
beyond the scope of this paper.}. This requirement leads to an upper
bound on the warmness $b$-parameter,

\begin{equation}\label{est}
\lim_{a\rightarrow 0}\frac{\Omega_{r}\left(a\right)}{\Omega_{w}\left(a\right)}>1 \qquad\Rightarrow\qquad b^{2} < \frac{\Omega_{r0}^{2}}{\Omega_{w0}^{2}-\Omega_{r0}^{2}}.
\end{equation}
Note that in early times, radiation dominates over baryons which decay
as $a^{-3}$. For this reason we do not take them into account in
Eq.~(\ref{est}). Since the present-day values are
$\Omega_{r0}\sim 10^{-4}$  and $\Omega_{w0}\sim 10^{-1}$, we expect
that $\,b^{2}\lesssim 10^{-6}$, which corresponds to a DM particle
velocity approximately equal to $300\,km/s$. Mathematically WDM
dominating over "standard''radiation means the absence of a real value for
$z_{eq}$,  that is the redshift at the point of radiation and matter
equilibrium. One can evaluate $z_{eq}$ from the relation
\begin{equation}
\Omega_{r}(z_{eq})=\Omega_{w}(z_{eq})+\Omega_{b}(z_{eq})\,.
\end{equation}
with the following solution,
\begin{eqnarray}
1+z_{eq}=\left(\frac{\Omega_{b0}\Omega_{r0}}{\Omega^2_{b0}-\Omega^2_{w0}}-\frac{\sqrt{(1+b^2)\Omega^2_{w0}\Omega^2_{r0}+b^2\Omega^2_{w0}[(1+b^2)\Omega^2_{b0}-\Omega^2_{w0}]}}{(1+b^2)\Omega^2_{b0}-\Omega^2_{w0}}\right)^{-1}.
\end{eqnarray}
\begin{figure*}[h!]
\resizebox{0.5\textwidth}{!}{%
  \includegraphics{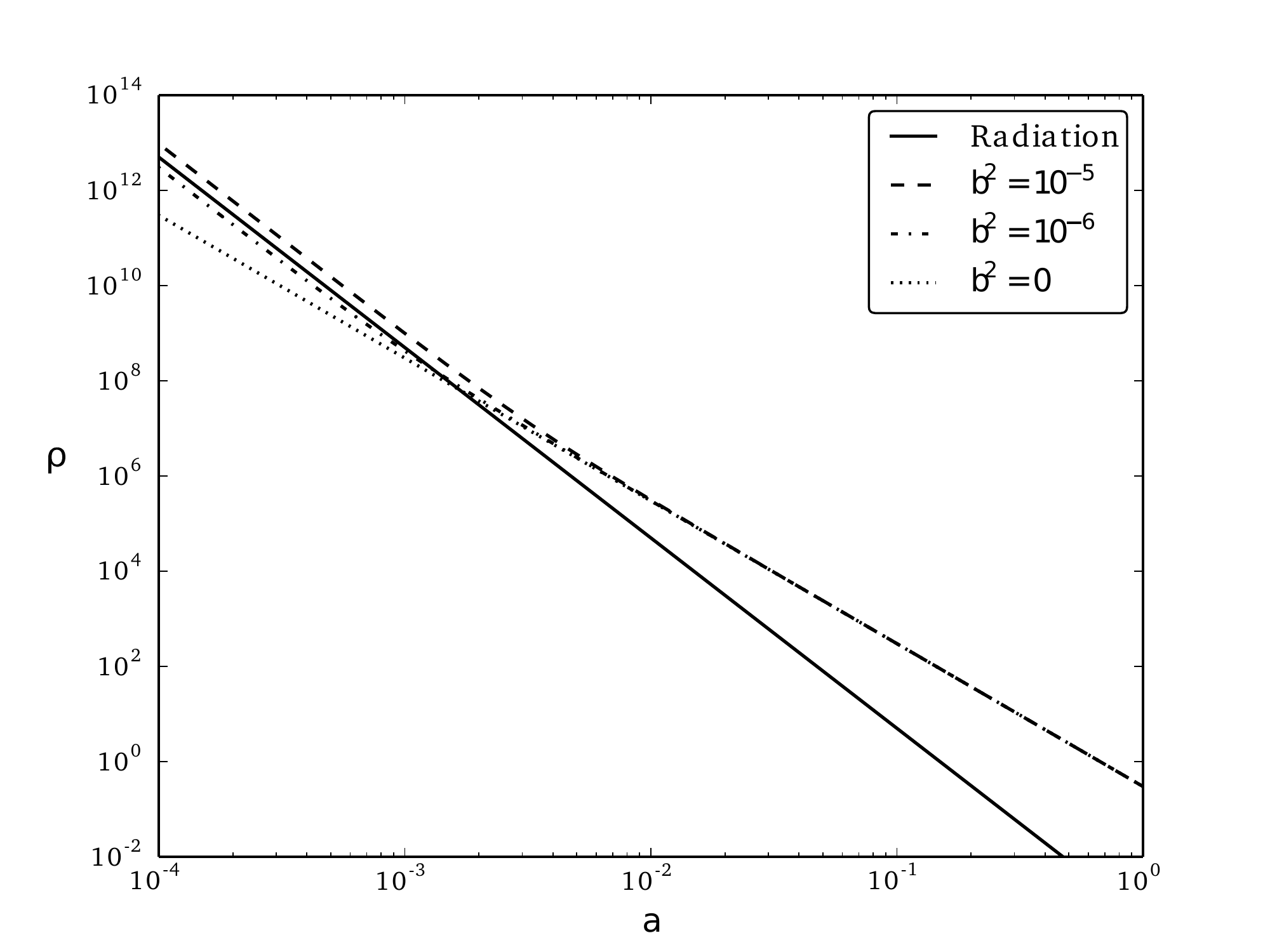}}
\resizebox{0.44\textwidth}{!}{%
  \includegraphics{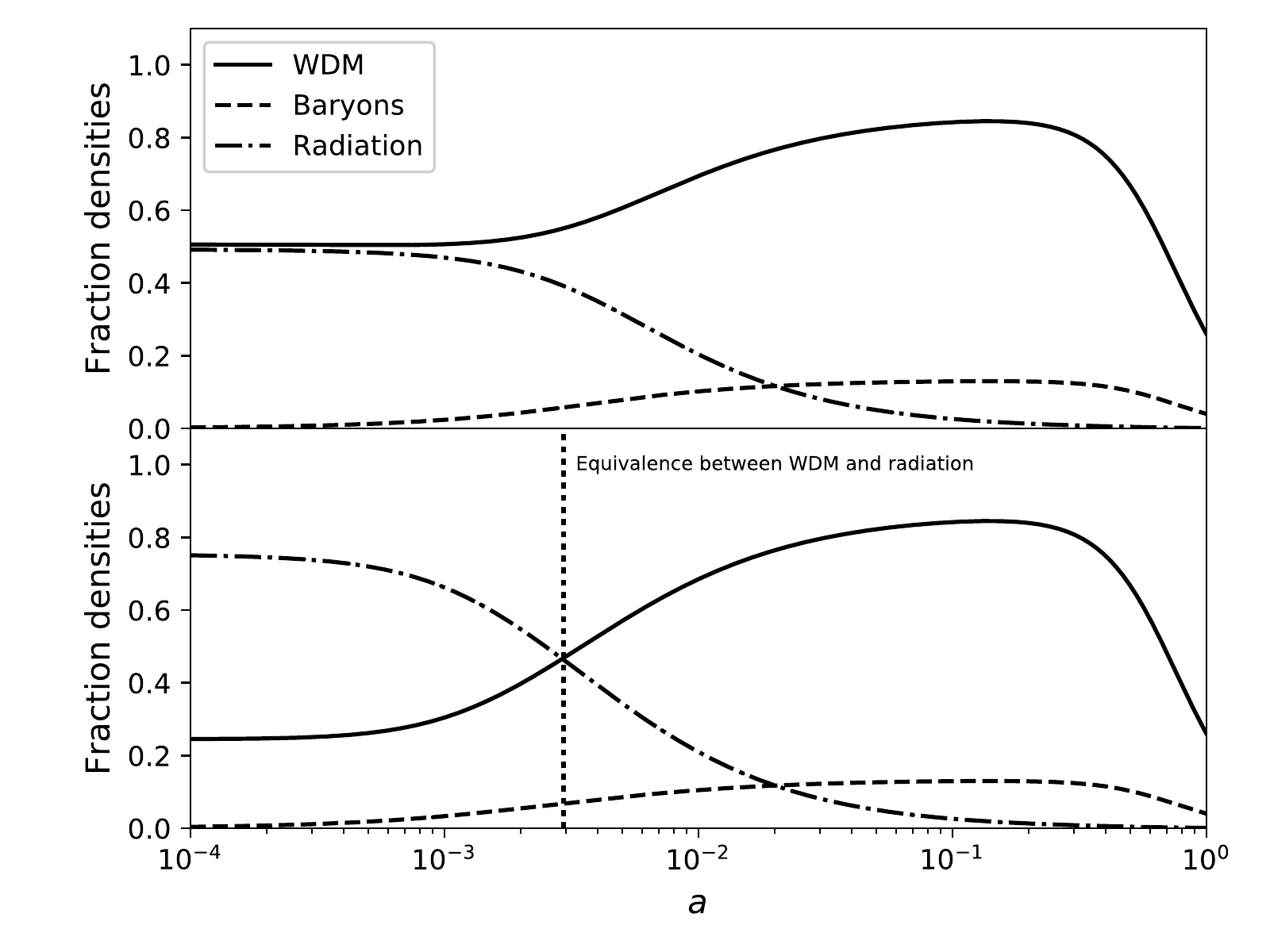}}
\caption{Left panel: Comparison between radiation density and  WDM density
for several values of $b^2$. For the values $b^2 \gtrsim 10^{-6}$
there is no radiation dominated era and WDM always dominates after
inflation. Right panel: Fractional abundances as functions of scale factor for
baryons, radiation and WDM for $b^2=10^{-5}$ (top) and
$b^2=10^{-6}$ (bottom).}
\label{fig:1}       
\end{figure*}

The early domination of WDM is shown in Fig.~\ref{fig:1}. The left panel of the
Fig.~\ref{fig:1} shows the densities of radiation and WDM for different
values of parameter $b$. We can see that for $b^2$-values higher than
$\sim 10^{-6}$ there is no radiation-dominated era. After inflation the
universe is always dominated by WDM. Moreover, for any value of
$b^2$-parameter smaller than $\sim 10^{-6}$, the equality between
WDM and radiation happens earlier compared to the CDM case. In the right panel of the 
Fig.~\ref{fig:1} one can see the plot for the scale factor dependence
of fractional abundances (i.e. $\Omega_i(a)/\Omega_T(a)$) for
radiation, baryons and WDM (here $\Omega_T(a)$ is the total
density parameter). The plot in the top panel corresponds to the case
$b^2=10^{-5}$ and clearly shows that WDM always dominates, while
in the bottom panel, for $b^2=10^{-6}$, we still have an epoch
dominated by standard radiation. In both cases the baryons contribution is
subdominant.
Consider now the structure formation process, which is strongly
dependent on the behavior of  WDM both at the background and
perturbative level. The dynamics of WDM perturbations has been
described in \cite{Fabris:2008qg}, so we can just write down the main
result for the dynamics of WDM perturbations. The energy and
momentum balance equation, in Fourier space for each $k$-mode
in flat universe  lead to following equations:
\begin{eqnarray}
\dot{\delta}_w+(1+w)\left(\theta_w+\frac{\dot{h}}{2}\right)+3{\cal{H}}\left(c^2_s-w\right)\delta_w+9{\cal{H}}^2\left(c^2_s-w\right)
 (1+w)\frac{\theta_w}{k^2} +  3{\cal{H}} \dot{w}\frac{\theta_w}{k^2}\ = 0 \,, \label{dpert} \\
 \dot{\theta}_w+{\cal{H}}(1-3c^2_s)\theta_w -\frac{k^2c^2_s}{1+w} \delta_w=0 \label{vpert} \,.
\end{eqnarray}
For the sake of convenience we used synchronous gauge and hence $h$ is the
trace of the scalar metric perturbations, $\delta_w\equiv\delta\rho_w/\rho_w$
is the WDM density contrast and $\theta_w$ is the velocity.  We have followed
conventions for metric signature and Fourier transform of \cite{Ma:1995ey}, and
the dot represents derivative with respect to conformal time $\eta$ and ${\cal{H}}=\dot{a}/a$.
Note that for $w=0$, CDM case is reproduced in the above equations.
The equations are written in the frame which is co-moving to the WDM
fluid and hence here was considered the rest-frame sound speed
$c^2_s$ \cite{Weller:2003hw,Gordon:2004ez}. We shall consider WDM as
adiabatic fluid. Actually, as far as we are dealing with thermal systems,
it is possible that some intrinsic non-adiabaticity traces could be present.
However, as a first approximation, we suppose that they are negligible.
Then one can use the relation
\begin{equation}
\delta p_w = c^2_s\delta\rho_w,
\end{equation}
where $c^2_s=\dot{p}_w/\dot{\rho}_w$.
Eqs.~(\ref{dpert}) and (\ref{vpert}) require analytical expression
for the rest-frame sound speed and the derivative of the state
parameter with respect to the conformal time. Using the background
quantities it is straightforward to obtain,
\begin{eqnarray}
\dot{w}=-\frac{{\cal{H}}}{3}\frac{a^2}{(a^2+b^2)}  \quad\mbox{ and }\quad c^2_s = w-\frac{\dot{w}}{3{\cal{H}}(1+w)} \,.
\end{eqnarray}
In order to solve the system it is necessary to fix initial conditions.
The WDM initial conditions can be implemented in the
super-horizon regime and deep into the radiation-dominated
epoch, i.e., $a\propto\eta$.  In the fluid description, for the early
radiation era, WDM case can be described by the following equations:
\begin{eqnarray}
\dot{\delta}_w+\frac{4}{3}\theta_w+\frac{2}{3}\dot{h}=0 \qquad\mbox{ and }\qquad \dot{\theta}_w -\frac{k^2}{4}\delta_w =0 \,.
\end{eqnarray}
By solving equation for $h$ in the super-horizon limit and
in the radiation era we arrive to the well-known
solution $h\propto (k\eta)^2$.
With the last solution we found, for the relevant limits, that
$\delta_w =-\frac{2}{3}C(k\eta)^2$
and
$\theta_w=-\frac{1}{18}Ck(k\eta)^3$ are appropriate initial conditions.
Of course, equations (\ref{dpert}) and (\ref{vpert}) are coupled with DE
via background solutions and with baryons and radiation both at the
background and perturbative level. One has to solve the complete
system in order to analyse the consequences of DM warmness for the
observables such as, e.g., CMB power spectrum, linear matter power
spectrum and the transfer function.

\section{Consequences of DM warmness via RRG}
\label{sec:3}

In addition to Eqs.~(\ref{dpert}) and (\ref{vpert}) we need also the
equations describing perturbations for baryons and radiation. These
equations can be found for example in \cite{Ma:1995ey} and we will
not repeat them here. To integrate the system including baryons,
radiation, WDM and cosmological constant, we modify the Boltzmann
CAMB code \cite{Lewis:1999bs}. The initial value
$\Omega_{b0}=0.0223 h^{-2}$ is chosen to provide the agreement
with Big Bang nucleosynthesis \cite{Pettini:2012ph}, while $\Omega_{r0}$
is taken to agree with CMB measurements \cite{Ade:2015xua}. The free
parameters related to WDM  are $H_{0}$, $\Omega_{w0}$ and $b$,
and in principle they have as priors $0<H_{0}<100$, $0<\Omega_{w0}<1$
and $0<b^2 \lesssim 10^{-6}$ to ensure a radiation dominated era. In this way we
consider a reduction of this WDM space of parameters by using
background  observational tests. Thus, in what follows we limit our
analysis to the values  of  WDM parameters such that they are in
$1\sigma$ CL (\textit{confidence level}) region of the  joint analysis
based on SNIa, BAO and $H(z)$ data. This shall help us to get more
realistic  and measurable  warmness effects that do not contradict
observations, at least at the background level.

\subsection{Background tests}
\label{sub:3.1}

The background  tests related to SNIa, BAO and $H_{0}$ are based on
the likelihood computed using the $\chi^{2}$ function,
\begin{equation}
\chi^{2}(\theta)=\Delta y (\theta)^T\mathbf{C}^{-1} \Delta y(\theta),
\end{equation}
where $\theta =(H_0,\Omega_{m0},b)$
and  $\Delta y(\theta) = y_i-y(x_i;\theta)$.
Here $y(x_{i};\theta)$  represents the theoretical predictions for a
given set of parameters, $y_i$ the data and $\mathbf{C}$  is the
covariance matrix. Note that, for convenience, 
the total matter density parameter $\Omega_{m0}=\Omega_{w0}+\Omega_{b0}$ was used here as a free parameter instead of $\Omega_{w0}$.

In order to perform the background statistical analysis it was used the numerical code CLASS \cite{Blas:2011rf} combined with the statistical code MontePython \cite{Audren:2012wb}. For the data set we have used the complete SNeIa data and correlation matrix from the JLA sample \cite{Betoule:2014frx}, $H_{0}$ is considered from \cite{Riess:2011yx},
and for BAO test we have used data from 6dFGS\cite{Beutler:2011hx}, SDSS \cite{Anderson:2013zyy}, BOSS CMASS \cite{Padmanabhan:2012hf} and WiggleZ survey \cite{Kazin:2014qga}.  The 6dFGS,
SDSS and BOSS CMASS data are mutually uncorrelated and also they are not correlated with WiggleZ data, however we must take into account
correlation beetween WiggleZ data points  given in \cite{Kazin:2014qga}. The set of 
free parameters $\theta$ can be divided in two parts: the cosmological free parameters $\Omega_{m0}$, $h$ and $b$; and the nuissance parameters 
$\alpha$, $\beta$, $M$ and $\Delta_{M}$, related to SNe Ia data. The results of the complete statistical analysis is presented in TABLE \ref{tab:1} and the 
contour curves are presented in FIG. \ref{fig:2}.
These results are in agreement with the previous results \cite{Fabris:2008qg,Fabris:2011am} but here we have updated
the results and  error was reduced due to the improved quality of observational data in recent years.
\begin{figure}[h!]
\resizebox{0.85\textwidth}{!}{%
  \includegraphics{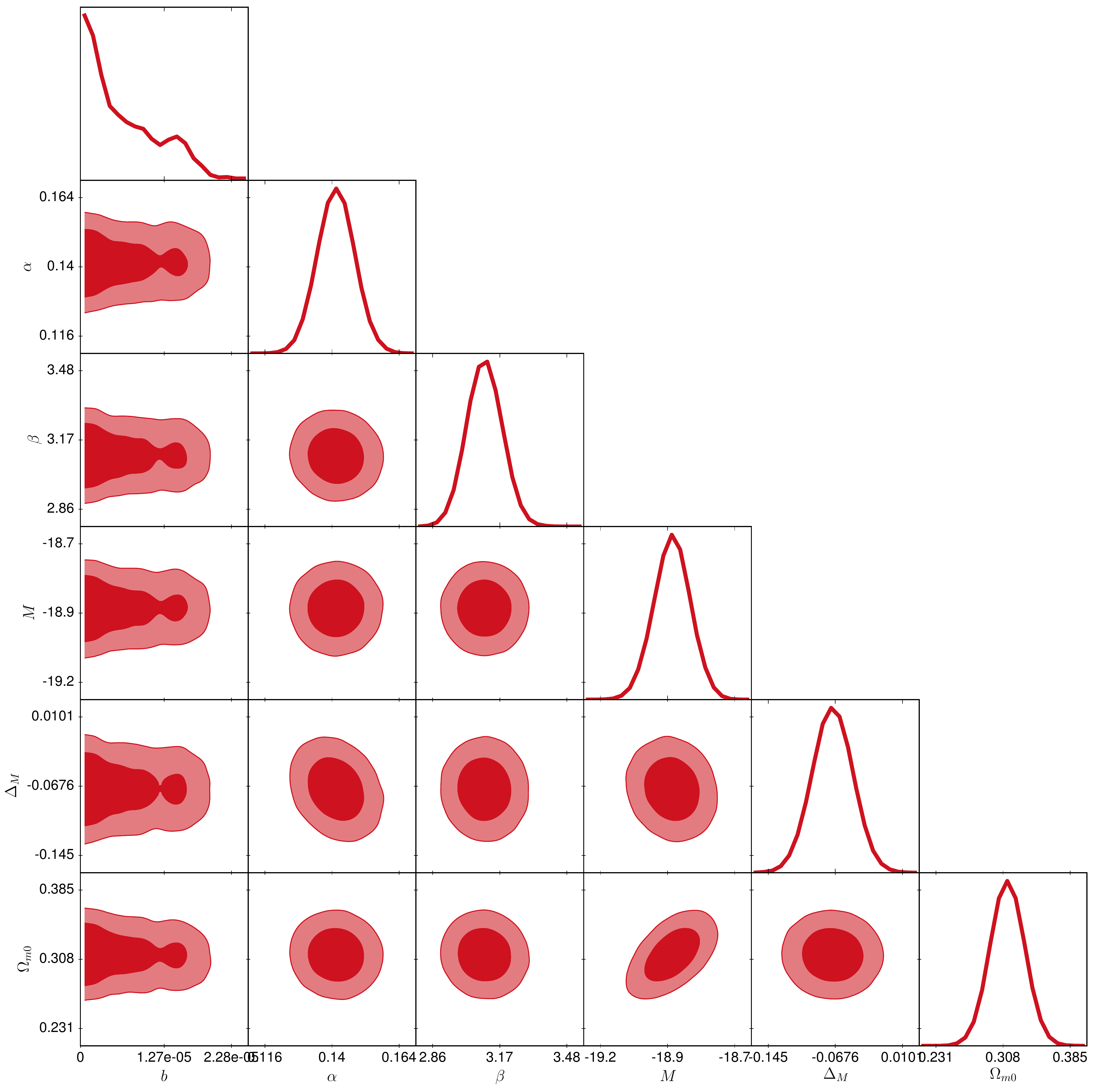}}
\caption{1$\sigma$ and 2$\sigma$ C.L. curves for the joint analysis using SNIa (JLA), $H_{0}$ and BAO data sets.}
\label{fig:2}       
\end{figure}

\begin{table}
\caption{Result of the statistical analysis for the joint analysis using SNIa (JLA), $H_{0}(z)$ and
BAO data sets. The best-fit result was obtained when $\chi_{min}^2=685.3$.}
\label{tab:1}       
\centering
\begin{tabular}{llll}
\hline\noalign{\smallskip}
Parameter & Best-fit & 95\% lower & 95\% upper \\
\noalign{\smallskip}\hline\noalign{\smallskip}
$h$ &$0.7386$ & $0.6923$ & $0.7833$ \\
$\Omega_{w0 }$ &$0.2607$ & $0.2237$ & $0.3025$ \\ 
$b$ &$6.425\times 10^{-6}$ & $0^{*}$ & $2.163\times 10^{-5}$ \\
$\alpha$ &$0.1417$ & $0.1277$ & $0.1545$ \\ 
$\beta$ &$3.094$ & $2.936$ & $3.265$ \\ 
$M$ &$-18.92$ & $-19.08$ & $-18.78$ \\ 
$\Delta_{M }$ &$-0.07238$ & $-0.1181$ & $-0.02448$ \\ 
$\Omega_{m0 }$ &$0.3107$ & $0.2737$ & $0.3525$ \\ 
\noalign{\smallskip}\hline
\end{tabular}
\end{table}

\subsection{Perturbative analysis}
\label{sub:3.2}

The reduced  space of parameters is in agreement with the
SNIa+BAO+$H(z)$ data, which we will use to study consequences
of the DM warmness in the two relevant observables, namely the
structure formation and CMB anisotopies. Before starting the
coprresponding consideration, let us illustrate the consequences
of the free-streaming of WDM in the matter perturbations.

Concerning the structure formation, a relevan quantity
is the total matter density contrast,

\begin{equation}\label{deltam}
\delta_{m}\equiv\frac{\delta\rho_{m}}{\rho_{m}}=\frac{\delta\rho_{w}+\delta\rho_{b}}{\rho_{w}+\rho_{b}}.
\end{equation}
By recalling that for each component $\delta\rho_{x}=\rho_{x}\,\delta_{x}$,
one can write analytic expression for the total matter density contrast in the
RRG-based model,
\begin{equation}\label{deltam1}
\delta_{m}=\frac{\tilde{\Omega}_{w}(a)\,\delta_{w}+\Omega_{b0}\,\delta_{b}}{\tilde{\Omega}_{w}(a)+\Omega_{b0}},\qquad\tilde{\Omega}_{w0}(a)=\Omega_{w0}\sqrt{\frac{1+b^{2}\,a^{-2}}{1+b^{2}}}\,.
\end{equation}
Let us note that in this case, different from $\Lambda$CDM, the total
matter density contrast depends on the scale factor. Furthermore, after
decoupling, the contribution of warm matter to total matter density
varies from $\sim 100\%$ for $a\ll1$ ($\delta_m \approx \delta_w$)
to $\sim 87\%$
when $a=1$ (i.e $\delta_m \approx 0.87\delta_w +0.13 \delta_b$)
while in $\Lambda$CDM the contribution is always constant and of
the order
$\sim 85\%$ (i.e $\delta_m \approx 0.85\delta_w +0.15\delta_b$).

The left panel of the FIG. \ref{fig:3} shows the total matter density contrast for different scales and for $b^2 = 10^{-14}$. In the top panel it is shown
$\delta_m$ for scale $k=2 h Mpc^{-1}$ and in the bottom panel it is shown $\delta_{m}$ for scale $k=5 h Mpc^{-1}$. In the first case the
difference with CDM case is minimal and $\sim 5\%$ at maximum. However, in the second case, this difference goes to $\sim 20\%$. These results indicate 
a strong suppression of the growth of matter
perturbations at the small scales, in contrast with the CDM case. Once again, one can see that the RRG enables
one to reproduce known features of WDM in a very economic way.

\begin{figure*}[h!]
\resizebox{0.45\textwidth}{!}{%
  \includegraphics{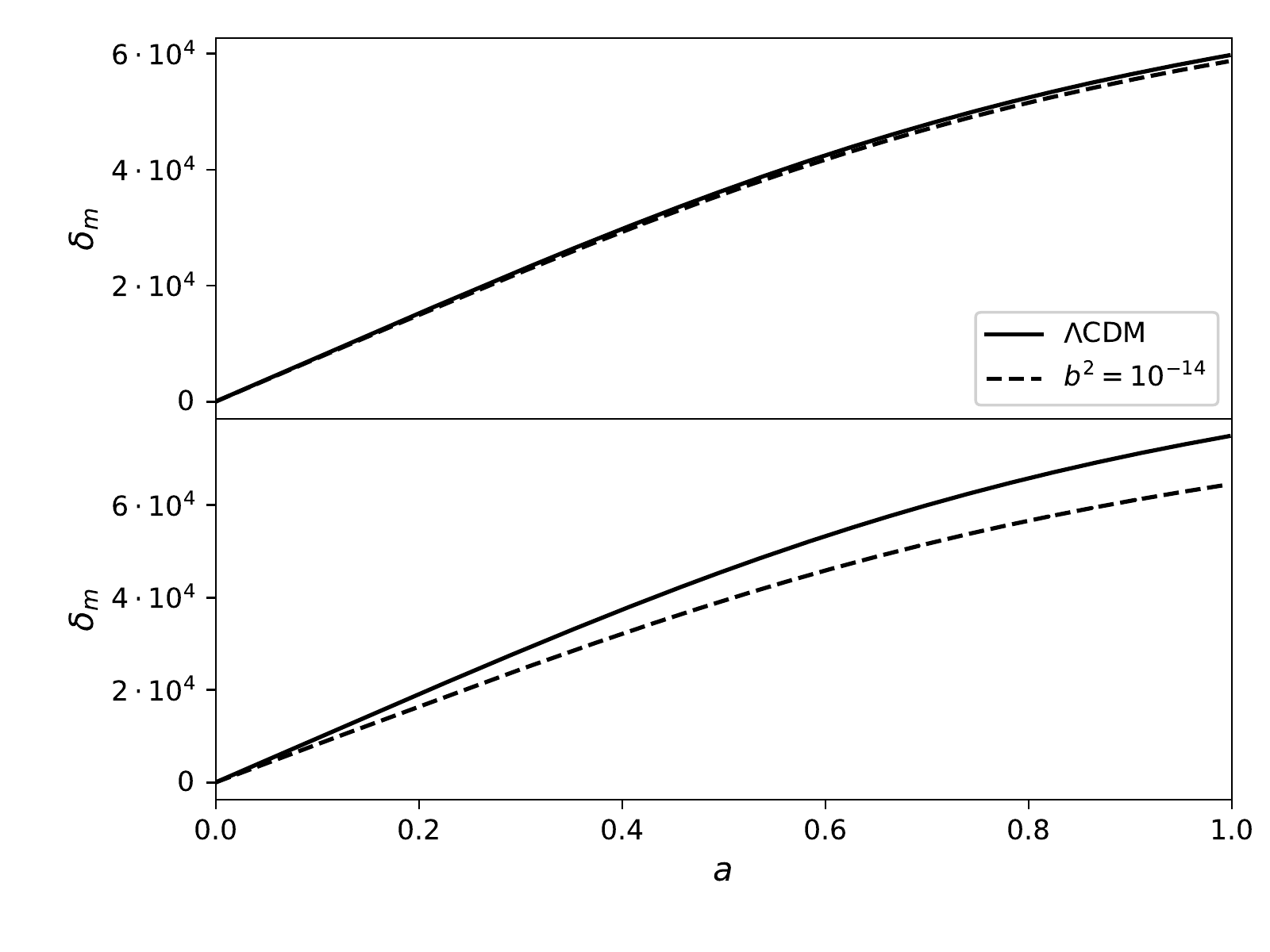}}
\resizebox{0.45\textwidth}{!}{%
  \includegraphics{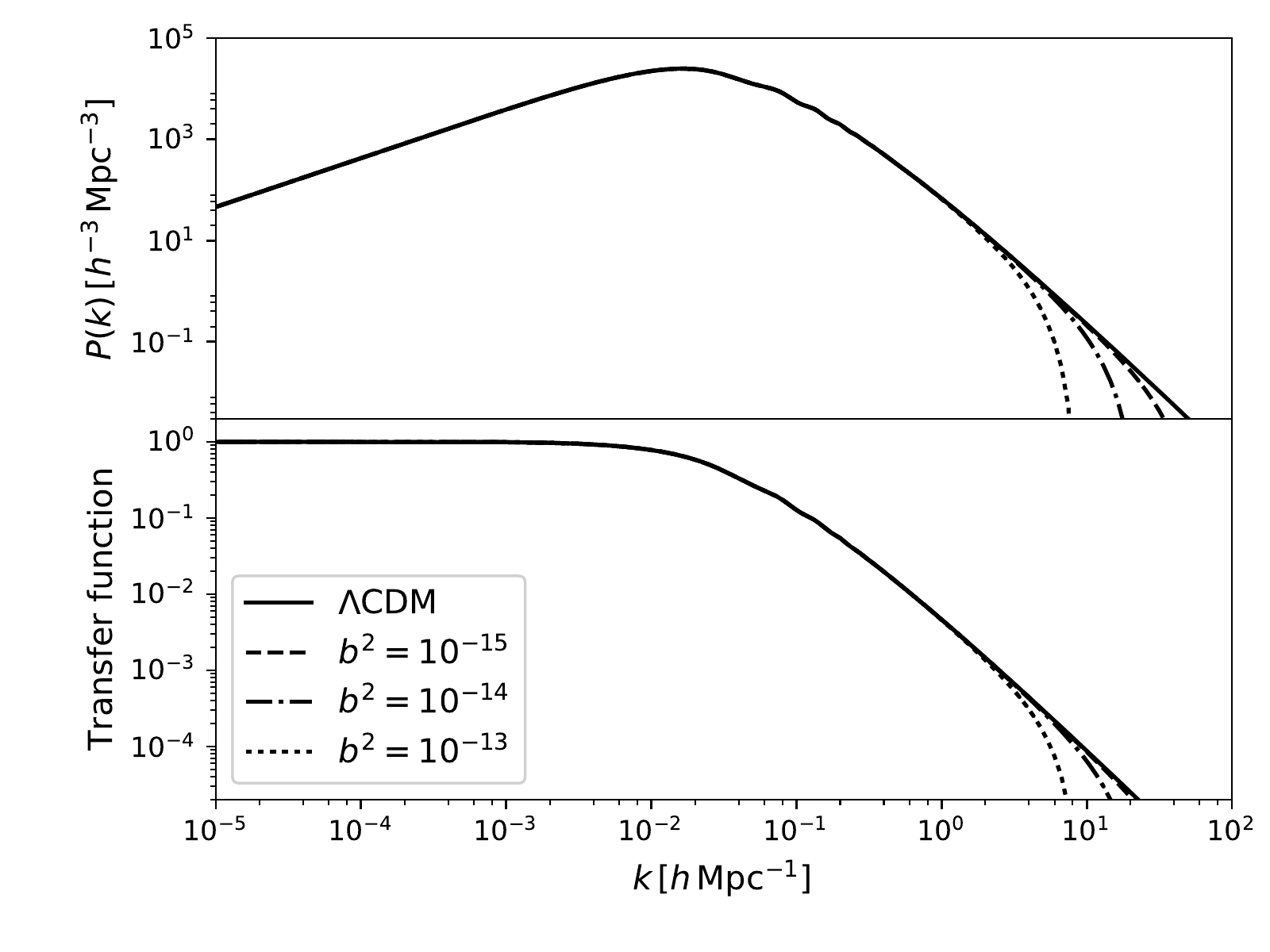}}
\caption{Top left panel: total matter overdensity  for $b=10^{-14}$ at scale $k=2 h Mpc^{-1}$. Maximum difference between WDM and CDM case
is $\sim 5\%$), bottom left panel represents the case for $k=5 h Mpc^{-1}$, where difference to the CDM case is $\sim 20\%$.  Top right
panel: linear matter power spectrum and bottom right panel: transfer function for different $b$-values. Note that suppression in small
scales is proportional to $b$ and is more evident in these quantities.}
\label{fig:3}       
\end{figure*}

One should expect that the suppression in the total matter density
contrast caused by DM warmness also appears in the linear matter
power spectrum and in its transfer function. The linear  matter
power spectrum is computed as $P(k)\propto k^{n_{s}}T(k)^{2}$,
where $n_{s}$ is the scalar spectral index and $T(k)$ is the
transfer function. The transfer function is defined as

\begin{equation}\label{tf}
T(k)\equiv\frac{\delta_{m}(k,z=0)\,\delta_{m}(0,z=0)}{\delta_{m}(k,z\rightarrow\infty)\,\delta_{m}(0,z\rightarrow\infty)}\,.
\end{equation}

At the next stage we use our modified CAMB code to compute the
linear matter power spectrum and transfer function. The results for
$b^{2}=10^{-13},10^{-14}$ and $10^{-15}$ are shown in the right panel of the FIG.
\ref{fig:3}.  The top right panel shows the linear matter power spectrum
while bottom right panel shows the transfer function for these cases.
From these plots one can conclude that at large scales there is no
much deviation from the $\Lambda$CDM result, while at the
small scales there is considerably lack of power proportional to
the value of warmness $b$ in relation to the $\Lambda$CDM.
This situation by itself is not new at all, it is regarded as one of
the main features of WDM models. However, it is remarkable
that one can reproduce it  by using the simple RRG description, in a model-independent way and without
any supposition about particle physics models.

One can wonder how CMB power spectrum is affected by the
suppression on matter overdensities  in small scales. The FIG.~\ref{fig:4} shows the CMB temperature power
spectrum for different $b$-values .
One can observe that even with the strong suppression in $P(k)$,
the CMB temperature power spectrum is not considerably affected
for $b^{2}\lesssim10^{-10}$. At large scales, when $l\lesssim 30$,
all curves  coincide. The differences only appear at the scales smaller
than $l \sim 30$. The most of the difference is at the intermediate
scales  $30 \lesssim l \lesssim 1300$. In order to quantify deviations
from $\Lambda$CDM we compute difference

\begin{eqnarray} \label{Dl}
\Delta D_l = D^{\Lambda CDM}_l-D^{\Lambda WDM}_l \,, \qquad D_l =\frac{l(l+1)C_l}{2\pi}
\end{eqnarray}
where $C_l$ represents the CMB temperature power spectrum.
Bottom panel of Fig.~\ref{fig:4} shows $\Delta D_l$. Notice
that the maximum difference is $ \lesssim 0.015\%$ and takes
place for $b^2=10^{-10}$. However, $\Delta D_l$ could be
slightly higher for the values of $b$ larger than $b^2 = 10^{-10}$.
Even though  large scales  $l \lesssim 30$ are not influenced by thermal velocities of
dark matter, the rest of the spectrum does. It is possible to show that some velocities
$\gtrsim 30 km/s$ ($b^2 \gtrsim 10^{-8} $) produce strong
distortions in the interval $30 \lesssim l \lesssim 1300$, such that
it would hardly fit the data.

\begin{figure}[h!]
\resizebox{0.5\textwidth}{!}{%
  \includegraphics{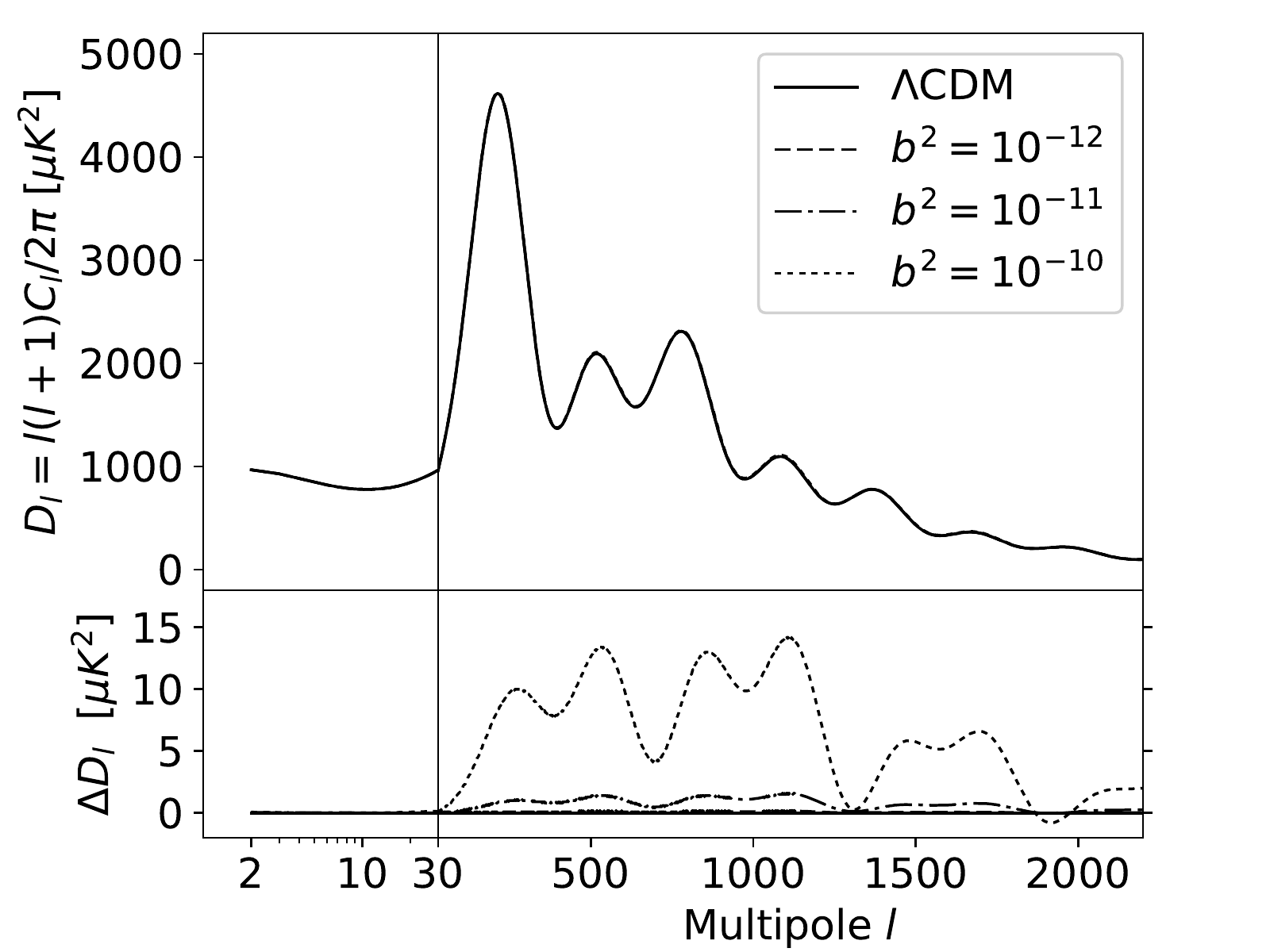}}
\caption{CMB temperature power spectrum for several values of $b$. Plots for $b^2=10^{-10}, 10^{-11}$ and $10^{-12}$ are shown in top panel. In bottom panel it is shown the difference $D_l$ defined in eq.(\ref{Dl}). Note that for $l\lesssim 30$ all curves are indistinguishable and differences appear after $l \sim 30$. For velocities of the order of $v\lesssim 3 km/s$, differences with CDM case is $\sim 0,015\%$ at maximum.}
\label{fig:4}       
\end{figure}

It is interesting to compare the RRG-based results with the ones 
which are based on different approaches. In the context of WDM, 
the effect of the free-streaming on matter distribution is quantified 
by  a relative function transfer  $\bar{T}(k)$ which is defined as

\begin{equation}
\bar{T}(k)\equiv\left[\frac{P_{\Lambda\rm{WDM}}(k)}{P_{\Lambda\rm{CDM}}(k)}\right]^{1/2},
\end{equation}
where $\,P_{\Lambda\rm{WDM}}\,$ and $\,P_{\Lambda\rm{CDM}}\,$
are linear matter power spectra for $\Lambda\rm{WDM}$ and 
$\Lambda\rm{CDM}$ cases, respectively. The function $\bar{T}(k)$ can 
be approximated by the following fitting expression \cite{Bode:2000gq},

\begin{eqnarray}
\label{barT}
\bar{T}(k)=[1+(\alpha k)^{2\nu}]^{-5/\nu},
\label{barTk}
\end{eqnarray}
where $\alpha$ and
$\nu$ are fitting parameters. 

For the sake of comparison, let us denote the relative function transfer 
computed via RRG by $\bar{T}_{RRG}(k)$. After computing  
$\bar{T}_{RRG}(k)$, we perform a fit for Eq. (\ref{barT}) and find 
parameters $\alpha$ and $\nu$ for different values of $b$. The results 
are summarized in three first entries of Table II. The results show that
RRG reproduces the relative transfer  function which is considered 
standard in the WDM framework with accuracy of $\lesssim 1\%$, which can be seen in FIG.~\ref{fig:5}.

In the left top panel of FIG.~\ref{fig:5}
we show  $\bar{T}_{RRG}(k)$ with $b^2 =10^{-15}$ and $\bar{T}(k)$ 
with $\alpha=0.0147$ and $\nu=1.12$, and in its bottom panel it is shown
the relative error between $\bar{T}_{RRG}(k)$ and
$\bar{T}(k)$. The same plots 
are shown in right panel of FIG. \ref{fig:5} for the case where the $\bar{T}_{RRG}(k)$ 
was computed with $b^2 =10^{-14}$ and $\bar{T}(k)$ was computed with
$\alpha=0.0242$ and $\nu=1.12$. Note that, for both cases, 
the relative error is $\lesssim 1\%$. 

\begin{figure*}[h!]
\centering
\resizebox{0.45\textwidth}{!}{%
  \includegraphics{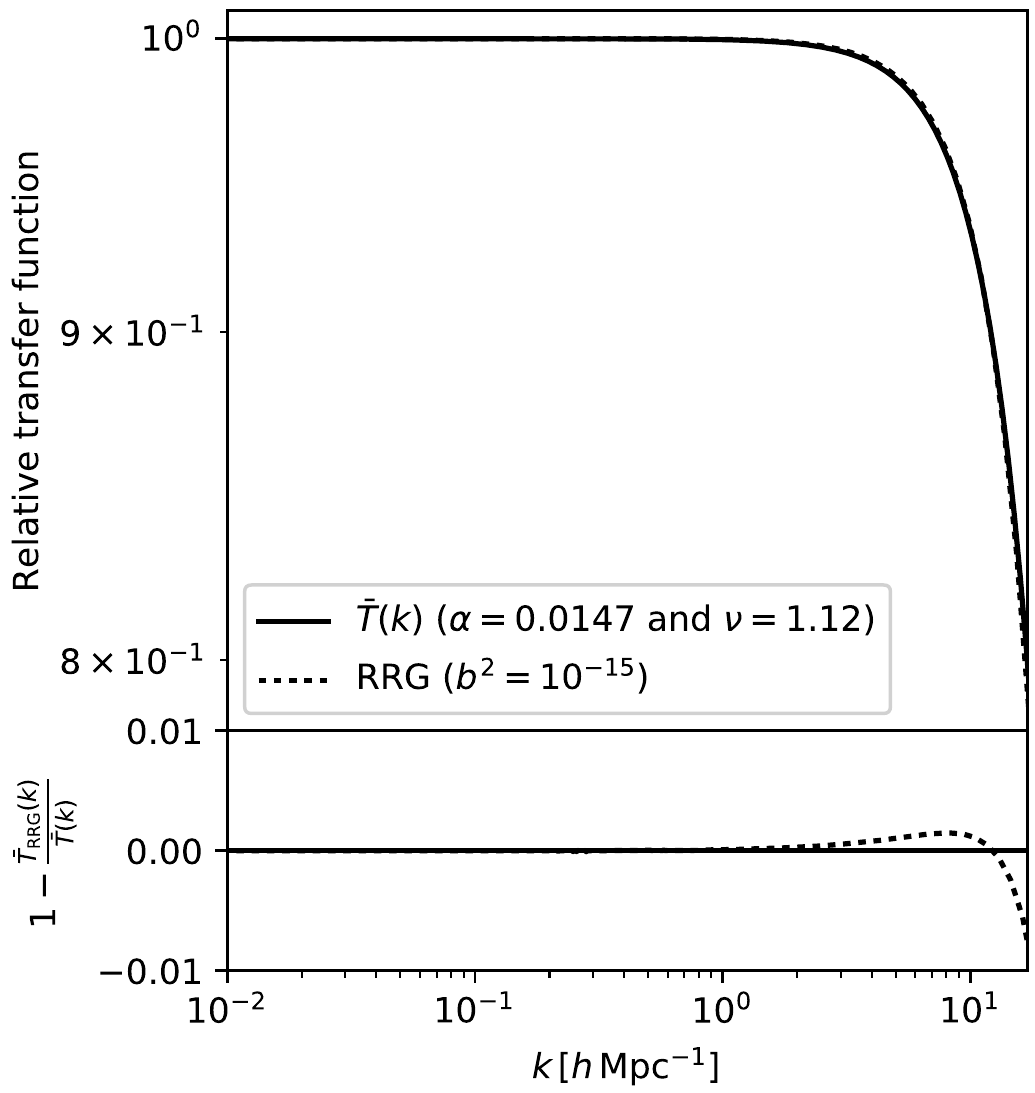}}
\qquad
\resizebox{0.45\textwidth}{!}{%
  \includegraphics{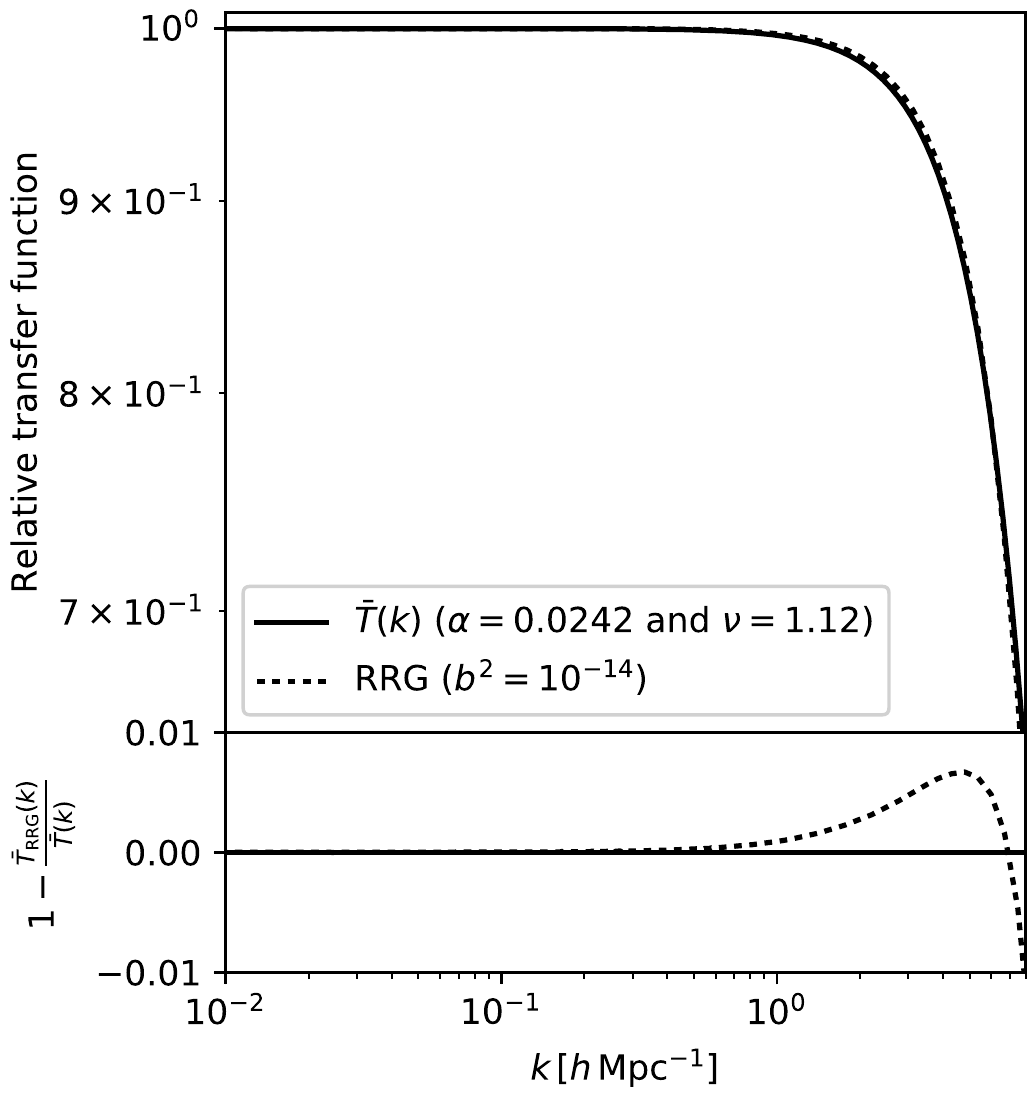}}
\caption{Top left: Plots for   $\bar{T}_{RRG}(k)$ with $b^2=10^{-15}$ and  $\bar{T}(k)$  with $\alpha =0.0147$ and $\nu=1.12$. Top right:
 $\bar{T}_{RRG}(k)$ with $b^2=10^{-14}$ and  $\bar{T}(k)$ with $\alpha =0.0242$ and $\nu=1.12$. Bottom left and right:
relative error for both cases.}
\label{fig:5}       
\end{figure*}

In what follows we shall consider a more detailed comparison between RRG approach and the well 
established particle physics candidate for WDM associated to thermal relics. Our comparison shall include 
some non-linear features.

\begin{table}
\caption{Result of the statistical analysis for the joint analysis using SNIa (JLA), $H_{0}(z)$ and
BAO data sets. The best-fit result was obtained when $\chi_{min}^2=685.3$.}
\label{tab:2}       
\centering
\begin{tabular}{lll}
\hline\noalign{\smallskip}
$b^2$ & $\alpha$ & $\nu$ \\
\noalign{\smallskip}\hline\noalign{\smallskip}
$10^{-10}$ & $2.450$ & $2.12$ \\
$10^{-11}$ & $0.510$ & $1.64$ \\
$10^{-12}$ & $0.350$ & $1.23$ \\
$10^{-13}$ & $ 0.092$ & $1.15$ \\
$10^{-14}$ & $0.028$ & $1.10$ \\
$10^{-15}$ & $0.005$ & $0.92$\\
\noalign{\smallskip}\hline
\end{tabular}
\end{table}

\section{Thermal relics via RRG}
\label{sec:4}

In the context of thermal relics it is well-known that there is a lower  bound\footnote{This limit comes from high redshift Lyman-alpha forest data \cite{Viel:2013apy}.} 
of $3.3\,keV$ for the WDM particle mass. Hence it would be interesting  to perform a comparison between thermal relics with such a bound and RRG approach.
For this reason, it is necessary first to find a equivalence between thermal relics mass scales and RRG $b$-parameter. Thus we recall that
for relics we have $\nu=1.12$  and the parameter $\alpha$, in units of $h^{-1}\rm{Mpc}$, is  related to the mass scale $m_{w}$ via 
\cite{Viel:2005qj,Bode:2000gq,Hansen:2001zv},
\begin{eqnarray}
\label{alfa}
\alpha = 0.049 \left(\frac{m_w}{1 keV}\right)^{-1.11}
\left(\frac{\Omega_{w0}}{0.25}\right)^{0.11}
\left(\frac{0.01 H_0}{0.7}\right)^{1.22}\,.
\end{eqnarray}

In order to obtain a complete association of the RRG parameter $b^{2}$ and the mass
of WDM particles in the thermal relics context, it was used the following particular set of values for the $b^{2}$,
\begin{equation}
b^{2}=\left(1\times 10^{-15},2\times 10^{-15},\,\ldots\, ,9\times 10^{-12},1\times 10^{-11}\right)\,.
\label{setb}
\end{equation}
For each point  it was defined a $\chi^{2}$ function,
\begin{equation}
\chi^{2}=\left(\bar{T}\left(k\right)^{th.}-\bar{T}\left(k\right)^{num.}\right)^{2}\,,
\label{chi2}
\end{equation}
where the $\bar{T}\left(k\right)^{th.}$ is given by the equation (\ref{barTk}) and the $\bar{T}\left(k\right)^{num.}$ is obtained with the 
CAMB code for each value of $b^{2}$ in the set (\ref{setb}). Then we minimize the equation (\ref{chi2}) in order to find the best-fit $\alpha$-value  
for each $b^{2}$.

Using the equation (\ref{alfa}) we found, for each value of $b^{2}$, the corresponding mass $m_{w}$. This correspondence is shown in FIG.~\ref{fig:6}, 
where the dots indicate the best-fit value for $m_{w}$ found through the equation (\ref{chi2}) and the solid line is the linear regression in the loglog 
frame. This linear regression results in the following fit-formula,
\begin{equation}
\label{mass}
m_{w}=4.65\cdot 10^{-6}\,\left(b^{2}\right)^{-2/5} keV.
\end{equation}

By using eq.(\ref{mass}) we found that a mass scale of  $3.3\,keV$ in thermal relics is equivalent to 
$b^{2}=2.36 \times 10^{-15}\sim10^{-15}$ in RRG approach. This value for $b^2$ brings  difficulties in distinguishing between WDM and CDM scenarios both at background 
and linear level. At background this can be seen in the left panel of the FIG. \ref{fig:1}, where for $b^2 \lesssim 10^{-6}$ the expansion dynamics after 
matter-radiation equality is indistinguishable from CDM case and also, the best fit value for $\Omega_{m0}$ is almost the same as $\Lambda$CDM (see FIG. \ref{fig:2}). On
the other hand, at linear regime, CMB signal via RRG for $b^2\lesssim10^{-10}$  is completly indistinguishable of $\Lambda$CDM 
(see FIG. \ref{fig:4}) and linear matter power spectrum has the expected little suppreession in small scales (see FIG. \ref{fig:3}). In order to better observe such tiny differences, we can for example, recompute the matter power spectrum in the Fourier 
space in its dimensionless form, denoted by $\Delta^{2}\left(k\right)=\frac{k^3}{2\pi^2}P(k)$ . 

\begin{figure}[h!]
\resizebox{0.5\textwidth}{!}{%
  \includegraphics{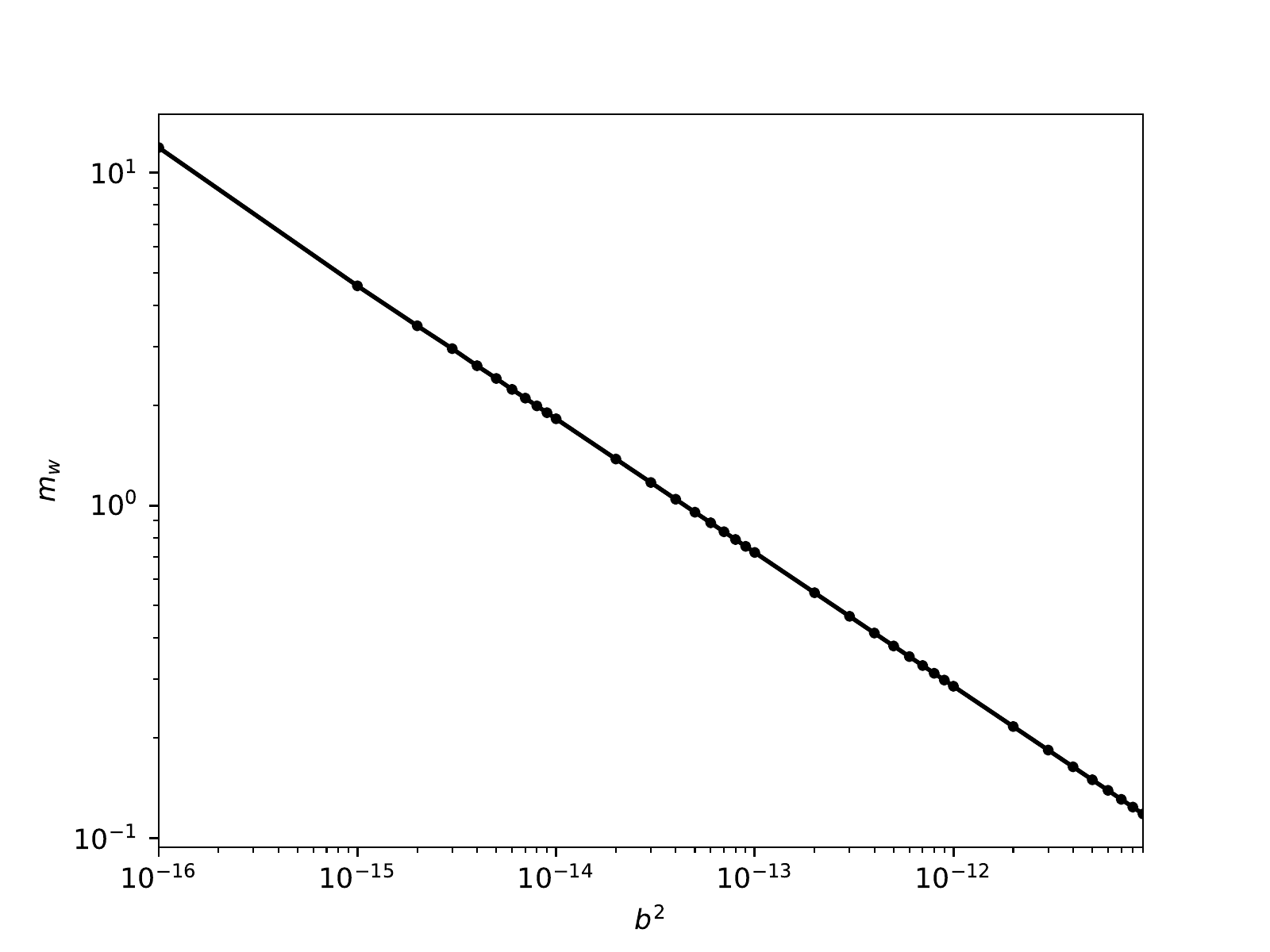}}
\caption{Fitting formula for the WDM mass particle (in $keV$) in terms of the RRG parameter $b^{2}$.}
\label{fig:6}       
\end{figure}

The left panel of the FIG. \ref{fig:7} shows $\Delta^2(k)$ at $z=0$ for both cases: standard treatment for 
thermal relics  with mass $3.3\,keV$ and thermal relics via RRG approach with  $b^{2}=10^{-15}$.
We can see that  CDM and WDM (thermal relics) are indistinguishable until  $\Delta^{2}\left(k\right) \sim 30$ where $\Delta^{2}\left(k\right)$
falls off too rapidly.  In fact, in the context of thermal-like candidates, this feature has been used
as justification to claim that albeit such a cut-off is still consistent with constraints based on Lyman-$\alpha$
forest data, for masses larger than 3.3 $keV$, the WDM appears no better than CDM in solving the small scale 
anomalies \cite{Schneider:2013wwa}. Of course, this affirmation needs to be better investigated in the context of others WDM candidates and, if possible, 
in a model-independent way. Thus, we believe that RRG can be very helpful in such direction.

On the other hand, we can also compute the time scale where the perturbations reach the non-linear regime.  
In the right panel of the FIG. \ref{fig:7} it is  shown the time (redshift) scale $z_{nl}$ in which the matter perturbation 
scale $R$ reachs the non-linear regime. $z_{nl}$ is the redshift where $\sigma^2(R)=1$, being $\sigma^2(R)$ the mass variance at the scale $R$. Again  both cases are considered: standard 
treatment for  thermal relics  with mass $3.3\,keV$ and thermal relics via RRG approach with  $b^{2}=10^{-15}$.
We can see that, in WDM thermal-like candidates context, the non-linearity is reached more recently than in $\Lambda$CDM for scales $R\lesssim1\, Mpc \,\,h^{-1}$.

Finally, by using $\Delta^{2}\left(k\right)$, $z_{nl}$ and a top-hat filter normalized with the results from Planck \cite{Ade:2015xua} 
it is possible to obtain the density of halos.  This quantity gives us the number of collapsed objects above a 
given mass $M$. Here, the mass function was computed using  the method of \cite{Sheth:2001dp}. In FIG.~\ref{fig:8}, the mass function obtained following  the RRG approach  is compared with the one computed 
for thermal  relics in the standard way and with CDM case. Clearly in WDM context, there is less collapsed objetcs than in 
$\Lambda$CDM. Furthermore, from FIGS. \ref{fig:7} and \ref{fig:8} it is possible to see the huge similarity between the RRG 
approach and the WDM thermal-like standard description also in the non-linear regime. In fact, the relative error in both cases 
is $\lesssim 1\%$.
\begin{figure*}[h!]
\centering
\resizebox{0.45\textwidth}{!}{%
  \includegraphics{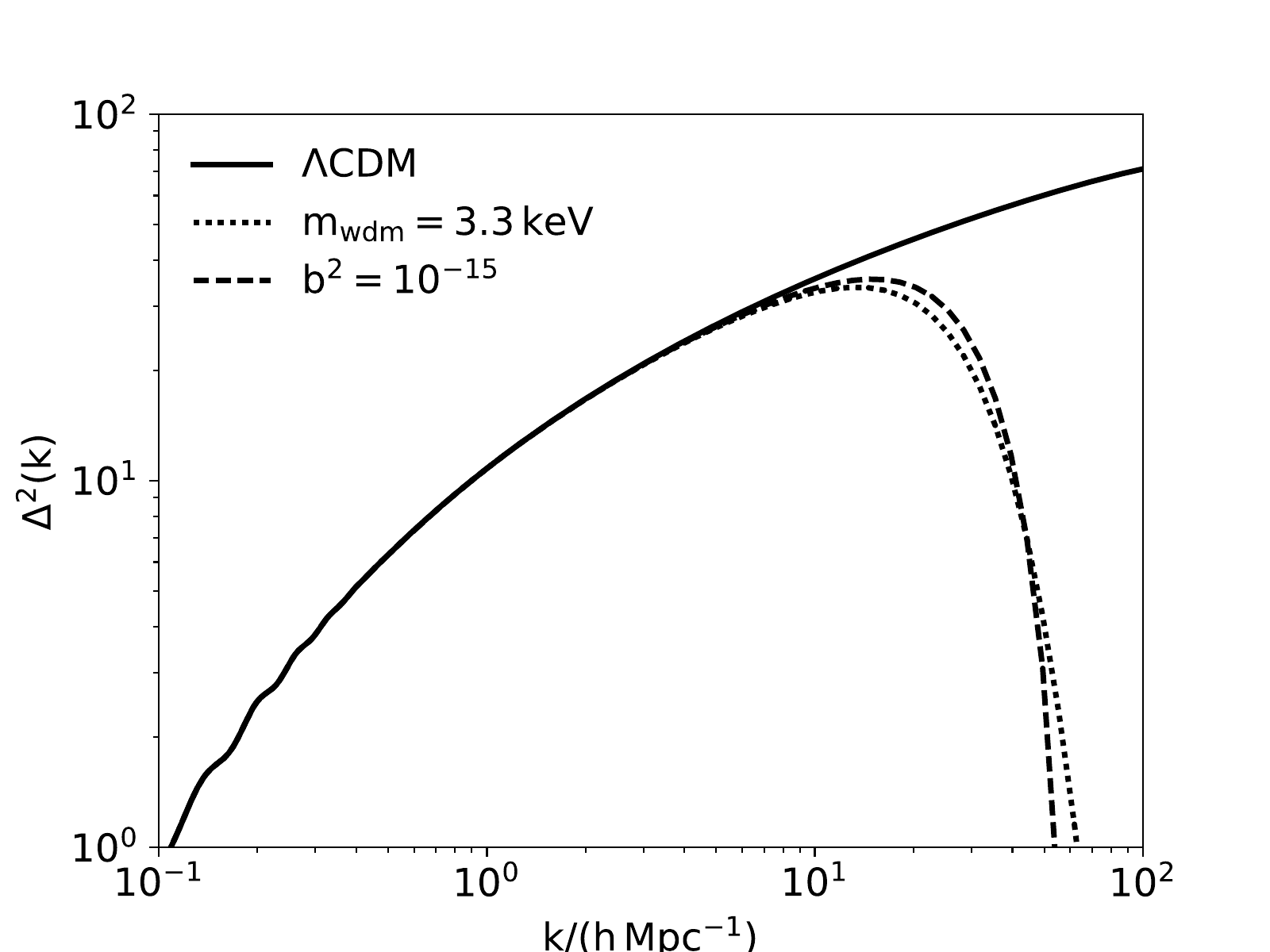}}
\qquad
\resizebox{0.45\textwidth}{!}{%
  \includegraphics{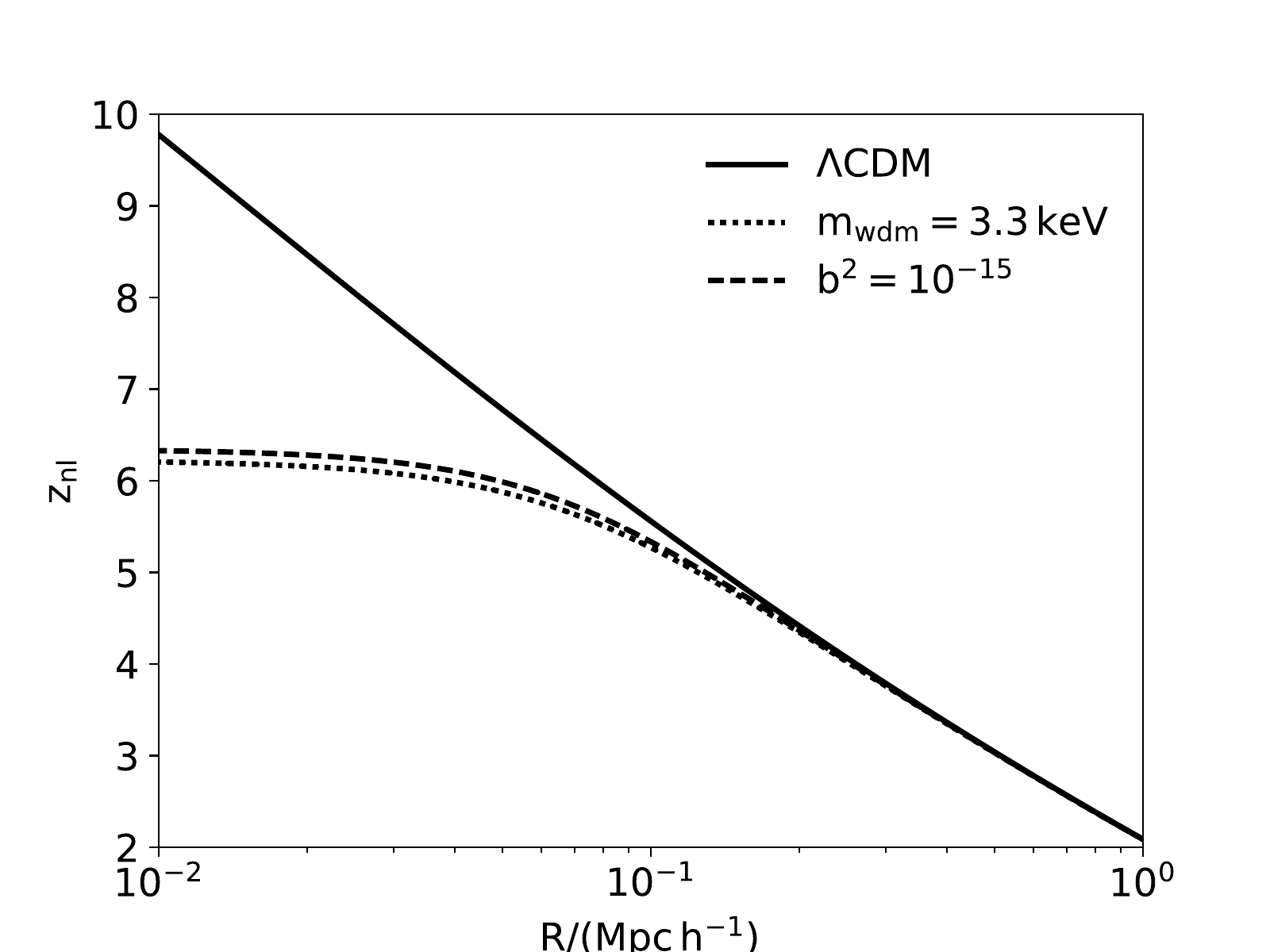}}
\caption{Left panel: Total matter perturbation in the Fourier space.
Right panel: Redshift where the matter perturbation reach the non-linear regime in function of the scale.
In both cases the solid line corresponds to $\Lambda$CDM model, the dashed line corresponds to the RRG approach and the dotted line corresponds to WDM case (thermal relics)
with mass $3.3\,keV$.}
\label{fig:7}       
\end{figure*}

Our results in this and in the previous section  indicate that  RRG approach is good
enough to capture important features  of WDM  in linear regime. Specially
the suppression on small scales structures and lack of power in
matter spectrum at such scales. Also, in the 
particular case of thermal relics, RRG reproduces with high precision, 
the potentialities and weakness of the candidate in both linear and 
non linear regime. Thus, RRG  approach could be considered as a complementary  
alternative approach to investigate  warm matter and specially
for understanding general behavior of the WDM scenario in a model-independent way.
\begin{figure}[h!]
\resizebox{0.5\textwidth}{!}{%
  \includegraphics{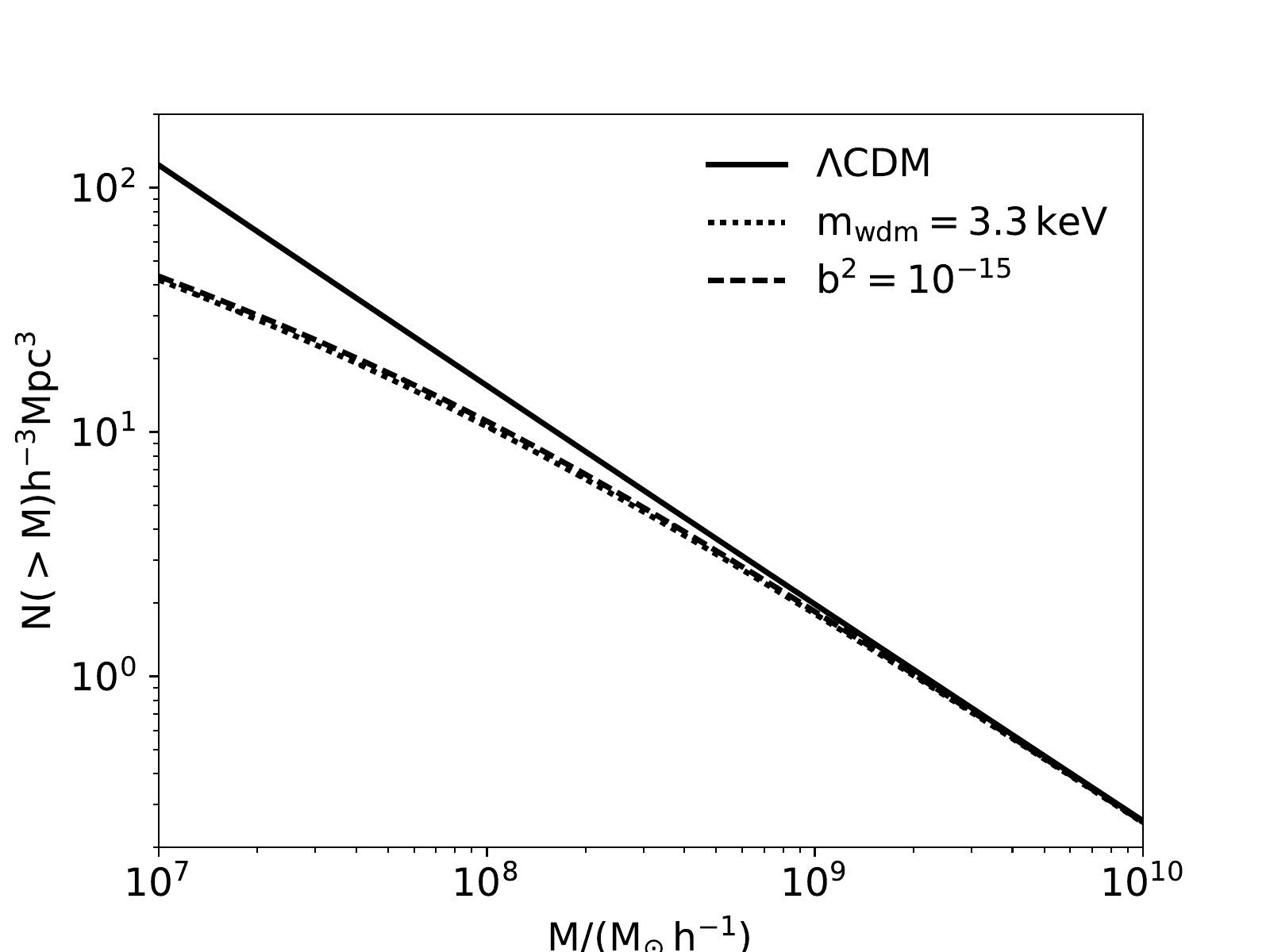}}
\caption{Mass function. The solid line corresponds to $\Lambda$CDM model, the dashed line corresponds to the RRG approach and the dotted 
line corresponds to standard approach  case with mass $3.3\, keV$.}
\label{fig:8}       
\end{figure}

\section{Discussion and conclusions}
\label{sec:5}

We have shown that the RRG approach enable one to
model WDM, which is treated as a gas of particles with
non-negligible thermal velocities. The simplifying aspect
of RRG is that all such velocities are taken to be equal.
The presence of warmness produce consequences on the
dynamics of the universe both for the background and
perturbations.

For the background the most important is that radiation era
may be smearing out for greater warmness. In this case the
universe is dominated by WDM up to the DE dominated era.
This scenario could have deep impact on the primordial
nucleosynthesis, reionization, recombination and other effects
in the early universe. This new non-standard scenario may be
deserving more careful and specific investigation, which we
leave for the future work. Our analysis here was limited by the
relatively small warmness, with $b^2 \lesssim 10^{-6}$.
In this case the radiation dominated era is still maintained, but
the radiation-matter equality takes place before than in the
CDM models. After the equality point, there are no serious
differences with expansion is dominated by cold matter, as it
is shown at the left panel of FIG.~\ref{fig:1}.

Instead of dealing with the full space of the WDM parameters,
we restrict consideration to a more reduced set. At the background 
level this is achieved by using recent data from SNIa, $H_{0}$ and 
BAO. As a result we arrive at reduced space which does not 
contradict more recent background observations at
$1\sigma$ CL, i.e  $b\in [0,2.1\times 10^{-5}]$ and $\Omega_{m0}=[0.27,0.35]$. Taking this into account one can expect 
that the quantification of imprint of warmness on observables should 
become more significant. Our analysis in this reduced  space of 
parameters shows that velocities which satisfy $\nu \lesssim 3\,km/s$ 
would agree with the CMB observations. This limit is essentially 
smaller than the bound for HDM, which may have velocities which are  
just two order of magnitude smaller than the speed of light. The 
velocities bound which were found here agree with the ones found 
earlier by other approaches 
\cite{Bond:1983hb,Hannestad:2000gt,Viel:2005qj,Bode:2000gq,Barkana:2001gr}.
This fact shows that, regardless of that the RRG is technically simple, 
it is a sufficiently reliable approach to probe new physics within the 
WDM approach. 

The consequences of a warmness of DM are is more evident in the
dynamics of cosmic perturbations. Since DM is supposed to be the 
main source of forming gravity potentials and overdensities, the 
impact of the DM warmness on structure formation and CMB 
anisotropies is evident. In the case of WDM thermal velocities cause 
free-streaming out from overdense regions, delaying and inhibiting the 
growth of fluctuations at certain scales. Another way to interpret this 
effect is by relating velocity to pressure. The non-negligible pressure 
of WDM, together with the radiation pressure, are resisting the 
gravitational compression and therefore suppress the power. This 
effect is stronger in small scales, as can be seen in Fig.~\ref{fig:3}.  
Furthermore, from Fig.~\ref{fig:4} one can conclude that  thermal 
velocities  do not affect considerably  the CMB temperature power 
spectrum for $b\lesssim 10^{-10}$. Indeed, the situation can 
be opposite for sufficiently large values of $b$.

As a next step we reproduced features of thermal relics by using RRG prescription. First it was necessary to 
find a relation between the mass scale in relics context and the $b^2$ parameter of RRG.
The $b^2 \sim 10^{-15}$ equivalent to the lower bound known for thermal relics ($3.3 keV$) brings the necessity to look  
more carefully matter perturbations. Thus, we computed the matter perturbations in the Fourier 
space $\Delta^2(k)$, the time scale where the perturbations reach the non-linear regime
($z_{nl}$) and the mass function. In all those cases, our results indicate that  RRG approach 
is good enough to capture important features  of WDM  even in non-linear regime.  Therefore, we  have proved that RRG is a reliable model, 
and can be considered as  a complementary, greatly simplified  alternative approach to investigate  
warm matter, in particular for understanding the behavior of the 
WDM  in a totally model-independent way.  

One can foresee the possibility of detailed investigations of the new 
Particle Physics candidates to WDM, which includes the comparison 
with model-independent RRG. In our opinion some aspects of this 
possibility would be quite interesting. For  example, let us mention 
a relation between $\bar{T}_{RRG}(k)$ and WDM candidate models, 
a more complete and comprehensive exploration of the space of parameters via Markov
Chain Monte Carlo (MCMC) or verifying how RRG would work in 
the nonlinear regime of structure formation through 
numerical simulations and the possibility if  WDM  is really capable of solve small scale problems. The work on these aspects of the model is currently in progress.

\section*{Acknowledgments}
Authors are grateful to Winfried Zimdahl for useful discussions.
J.F. and R.M. wish to thank  CAPES, CNPq and FAPES
for partial financial support. I.Sh. acknowledges the partial
support from CNPq, FAPEMIG and ICTP.

\bibliographystyle{ieeetr}
\bibliography{WDMviaRRG}

\end{document}